\documentclass[prb,superscriptaddress,twocolumn,showpacs,preprintnumbers,amsmath,amssymb]{revtex4}

\usepackage{graphicx}
\usepackage{dcolumn}
\usepackage{bm}

\begin{document}

\title{Magnetodielectric Response of the Spin-Ice Dy$_2$Ti$_2$O$_7$}

\author{Masafumi~Saito}
\email{msaito@scphys.kyoto-u.ac.jp}
\affiliation{Department of Physics, Graduate School of Science, Kyoto University, Kyoto 606-8502, Japan}

\author{Ryuji~Higashinaka}
\affiliation{Department of Physics, Graduate School of Science, Kyoto University, Kyoto 606-8502, Japan}

\author{Yoshiteru~Maeno}
\affiliation{Department of Physics, Graduate School of Science, Kyoto University, Kyoto 606-8502, Japan}
\affiliation{International Innovation Center, Kyoto University, Kyoto 606-8501, Japan}

\date{\today}

\begin{abstract}
We report the magneto-dielectric response of single crystals of the spin-ice compound Dy$_2$Ti$_2$O$_7$ down to 0.26 K. 
The dielectric constant under zero magnetic field exhibits a clear decrease reflecting the development of the local two-spins-in, two-spins-out structure below about 1.2 K. 
Both the real and imaginary parts of the dielectric constant under magnetic fields sensitively respond to various changes in the spin structures. 
We found that the real part can be described in terms of local spin correlations among the moments of tetrahedra, rather than among individual Dy$^{3+}$  moments.  
Using the peaks in the imaginary part, we have constructed a precise field-temperature phase diagram in the [111] field direction. 
We thus demonstrate that the magneto-dielectric response can be a high-sensitivity local probe of the spin state of geometrically frustrated systems.
\end{abstract}


\maketitle

\section{Introduction}

Geometrically frustrated spin systems are now clearly recognized as a class of magnets exhibiting unusual behavior such as residual entropy, spin chirality, and quantum spin-liquid state~\cite{Ramirez1}. 
For the experimental investigation of frustrated mechanisms leading to such behavior, it is crucial to probe the local changes in the spin configuration not always probed readily by conventional bulk methods. 

\begin{figure}[b]
    \begin{center}
\includegraphics[width=0.8\linewidth,clip]{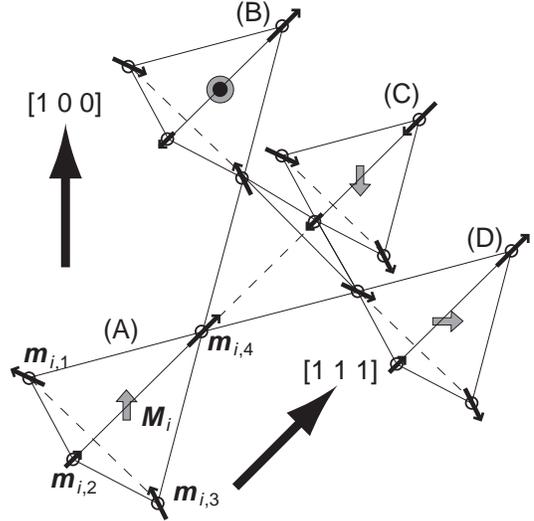}
    \end{center}
\caption{Network of tetrahedra in the pyrochlore lattice. Black arrows at vertices represent the magnetic moment ${\bm m}_i$ of Dy$^{3+}$. Gray arrows and gray circle represent the total moment ${\bm M}_i$ for a tetrahedron, defined by ${\bm M}_i \equiv \sum_{j=1}^{4} {\bm m}_{i,j}$. Tetrahedra (B)-(D) constitute three of the six nearest-neighbors for (A).}
\label{fig:Tetra}
\end{figure}

Recently, the strong coupling between dielectric and magnetic properties of titanates~\cite{Katsufuji1} and manganites~\cite{Kimura} attracted renewed research interest.
A phenomenological expression for the magneto-dielectric response, which takes into account of local spin correlations, has been proposed~\cite{Katsufuji1}.
This motivated us to investigate how the dielectric constant responds to a variety of changes in the spin configuration in the spin-ice magnet, an archetypal system with geometrical spin frustration. 
In this study, we use the spin-ice compound Dy$_2$Ti$_2$O$_7$ in the pyrochlore structure~\cite{Bramwell}, in which Dy$^{3+}$ ions constitute a three-dimensional network of corner-shared tetrahedra, so called the pyrochlore lattice shown in Fig.~\ref{fig:Tetra}.
Because of crystal field splitting of the $J = 15/2$ multiplets, the ground-state of the Dy$^{3+}$ ion is a doublet with the magnetic moment exhibiting strong ``Ising anisotropy" along the local $<$111$>$ direction, which point to the center of a tetrahedron from a vertex. 
Dipolar interaction dominates the nearest-neighbor interaction and is ferromagnetic, although the antiferromagnetic exchange interaction is also substantial. 
Thus the spin configuration at low temperature consists of two of the four spins in a tetrahedron pointing inwards, and the others outwards (the ``2-in, 2-out" state). 
Because of the statistical equivalence to the problem of hydrogen configuration in water ice, the 2-in, 2-out constraint is called the "ice rule"~\cite{Bramwell}.
Since there are a macroscopic number of degenerate states under the 2-in, 2-out ice-rule based configuration, the system exhibits the so-called spin ice state~\cite{Ramirez2,Fukazawa,Snyder1} with a residual entropy.

Very recently, Katsufuji and Takagi reported the magnetocapacitance of the spin-ice system Ho$_2$Ti$_2$O$_7$~\cite{Katsufuji2}, but only for temperatures above 1.8 K. 
Here we report the magneto-dielectric response to temperatures low enough to probe the changes in the spin state associated with geometrical frustration.
The results reveal that both the real part $\varepsilon^{\prime}(T, B)$ and the imaginary part $\varepsilon^{\prime\prime}(T, B)$ of the dielectric constant can be powerful high-sensitivity probes for frustrated magnets.

In the [100] field direction, in which all four spins in each tetrahedron are equally affected by the magnetic field, a crossover occurs at about 0.5 T to a state without spin degeneracy at low temperatures, in which all ${\bm M}_i$s point along [100] (Fig.~\ref{fig:Tetra}). 
In the [111] field direction, in which one of the four spins is exactly parallel to the magnetic field, the 2-in, 2-out state becomes unstable for fields greater than about 1 T and changes to the ``1-in, 3-out state" in which all ${\bm M}_i$s have a positive component along the field. 
The pyrochlore lattice can be viewed as a stacking of ``pyrochlore slabs" consisting of triangular-lattice and Kagome-lattice planes normal to the [111] direction. 
It was recently found that in an intermediate field range, the field-parallel spins on the triangular lattice align along the field, whereas the other spins on the Kagome lattice maintain frustration and associated residual entropy because of the overall 2-in, 2-out ice-rule constraint (the Kagome-ice state)~\cite{Matsuhira, Hiroi1, Higashinaka1}. 
Because of such variety of frustrated spin structures which can be controlled by magnetic fields in particular directions, the spin-ice magnet is ideal to investigate how the magneto-dielectric response can probe the local spin structures. 

\section{Experiment}
We used single crystals of Dy$_2$Ti$_2$O$_7$ grown with an infrared floating-zone furnace (NEC Machinery, model SC-E15HD)~\cite{Fukazawa} and cut to a typical size of 3 $\times$ 2 $\times$  0.7 mm$^3$ with the crystalline axis [100] or [111] normal to the wide plane of the sample. 
We measured the magneto-dielectric response as well as magnetostriction with a capacitance cell designed for a standard three-terminal method. 
We used a capacitance bridge (Andeen-Hagerling, model 2500A) with a fixed measurement frequency of $f = 1$~kHz. 
The capacitance cell was cooled down to 0.26 K using a $^3$He refrigerator (Oxford Instruments, model Heliox VL) under magnetic fields of up to 7 T. 
The thermal contact of the sample was made by gold films sputtered on the top and bottom sample surfaces in direct contact with the copper pole plates, which are connected to a thermally-anchored copper coaxial cable.
The gold coating was needed also to eliminate vacuum gap between the sample and the pole plates. 

The measured capacitance $C$ was about 10 pF at room temperature, increased by 20 \% with decreasing temperature, and peaked at about 20 K. 
The measured dielectric loss $A$ was of the order of 10$^{-1}$ nS at room temperature and decreased by up to a factor of ten at low temperatures. 
In all the Dy$_2$Ti$_2$O$_7$ crystals we investigated, we reproducibly observed a strong and sharp peak in $A$ at about 150 K with the width of only about 20 K. 
Although it is similar to the energy scale of the crystalline field excitation level of the Dy$^{3+}$ ion, the sharpness of the peak in $A$ suggests that it is not simply due to thermal activation of Dy$^{3+}$ moments.
Its origin is not unknown at present. 
The imaginary part of the dielectric response is evaluated from $A$ and $C$ by  $\varepsilon^{\prime\prime} = \varepsilon^{\prime}\tan\delta$ $=\varepsilon^\prime A / (2 \pi f C)$. 
In the data presented below, the magnetic field has been corrected for the demagnetization factor based on the magnetization data at respective temperatures~\cite{Fukazawa,Sakakibara}; the field is expressed in the internal field $B$, rather than in the external field $H$. 

\begin{figure}[b]
    \begin{center}
\includegraphics[width=0.9\linewidth,clip]{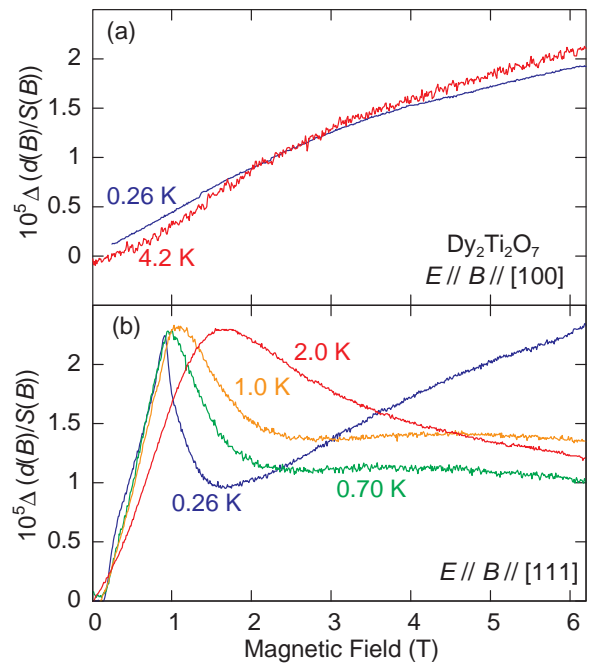}
    \end{center}
\caption{(Color online) Magnetostriction of Dy$_2$Ti$_2$O$_7$ for the (a) [100] and (b) [111] directions. $d(B)$ is the sample thickness and $S(B)$ is the sample area under the magnetic field $B$.
Curves in (b) are offset from each other for clarity.}
\label{fig:MagSt}
\end{figure}

Since the magnetocapacitance contains information not only of the dielectric constant but also of changes in the sample thickness and area, it is important to examine the contribution from the magnetostriction.  
For this purpose, we used the same capacitance cell but glued on the lower pole-plate a sample crystal coated with sputtered gold film on all its surfaces. 
In addition, we placed three glass posts, each 1 $\times$ 1 $\times$ 0.84 mm$^3$ in size, between the pole plates so that there is a spacing of ca. 100 $\mu$m between the crystal and the upper pole plate. 
From the measurement of the capacitance with and without the sample crystal under magnetic fields, we extracted the changes in the sample shape $\Delta(d(B)/S(B))$, where $d$ is the thickness and $S$ is the area of the sample surface. 
The results for the [100] and [111] field directions are shown in Fig.~\ref{fig:MagSt}. 
As shown below, the magnetostriction is more than an order of magnitude smaller than the magneto-dielectric response.
For the [100] field direction shown in Fig.~\ref{fig:MagSt}(a), for which all the spins respond equally to the Zeeman energy of the external field, the magnetostriction increases monotonically with field. 

\begin{figure}[t]
    \begin{center}
\includegraphics[width=0.9\linewidth,clip]{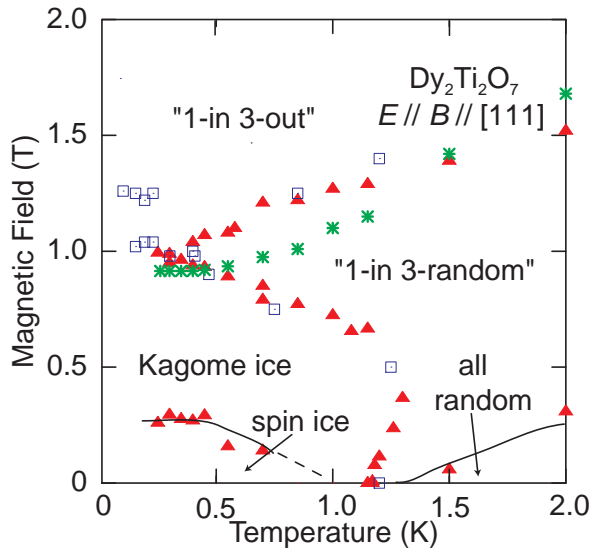}
    \end{center}
\caption{(Color online) Field-temperature phase diagram of Dy$_2$Ti$_2$O$_7$ for the [111] field direction. Closed triangles indicate the peak position in the dielectric loss, open squares the peak in the specific heat~\cite{Higashinaka2}, and asterisks the peak in the magnetostriction. The lines are guides to the eye.}
\label{fig:Phase}
\end{figure}

By contrast, we observed a clear peak for the [111] field direction as shown in Fig.~\ref{fig:MagSt}(b). 
Compared with the specific-heat peak position shown in Fig.~\ref{fig:Phase}, we conclude that the magnetostriction peak reflects the change in the spin configuration to the ``1-in, 3-out" polarized state.
Corresponding to this change becoming a first-order transition below 0.4 K~\cite{Sakakibara}, the magnetostriction peak is very sharp at 0.26 K. 
Moreover, a shoulder feature at 0.4 T corresponds well to the change from the spin ice to the Kagome ice. 
Nevertheless, other spin-configuration boundaries in the phase diagram cannot be probed by magnetostriction within the present precision. 
It is interesting to note that the height of the peak remains essentially unchanged with temperature. 
The results here are semi-quantitatively consistent with those reported for polycrystalline Dy$_2$Ti$_2$O$_7$ at 1.7 and 4.2 K, for which the longitudinal magnetostriction is substantially bigger than the transverse one~\cite{Mamsurova}. 

\begin{figure}[b]
    \begin{center}
\includegraphics[width=0.9\linewidth,clip]{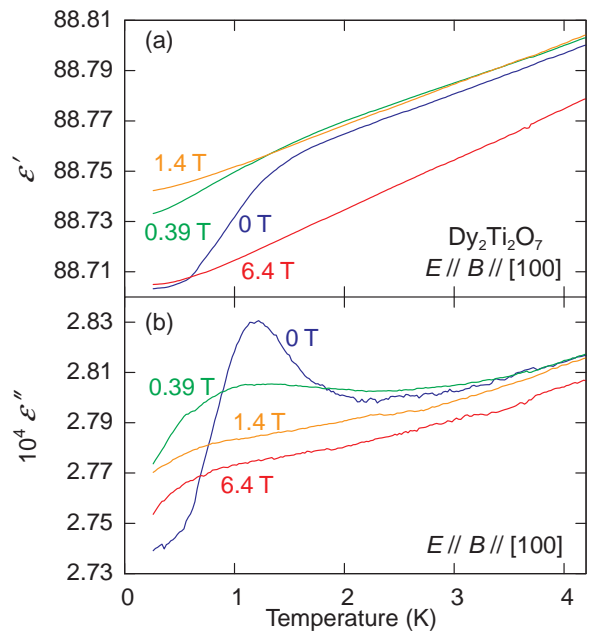}
    \end{center}
\caption{(Color online) Temperature dependence of (a) the real part $\varepsilon^\prime$ and (b) the imaginary part $\varepsilon^{\prime\prime}$ of the dielectric constant of Dy$_2$Ti$_2$O$_7$ for magnetic fields along the [100] directions. The data were obtained on warming after cooling under the presence of the magnetic field.}
\label{fig:EvsT-100}
\end{figure}

\begin{figure}[t]
    \begin{center}
\includegraphics[width=0.9\linewidth,clip]{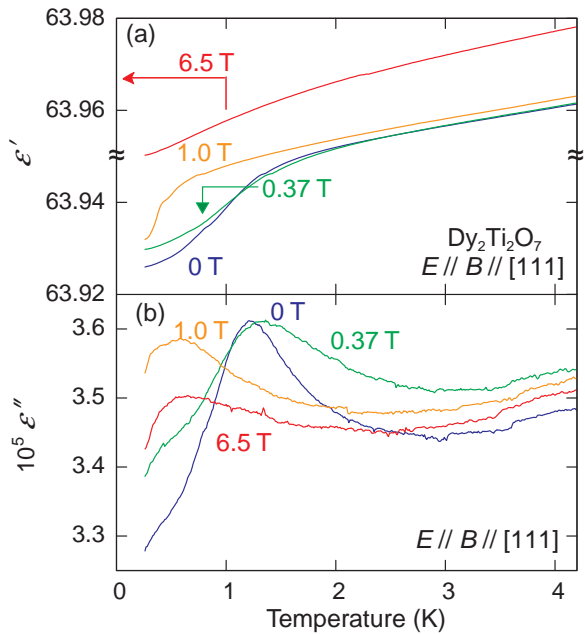}
    \end{center}
\caption{(Color online) Temperature dependence of (a) the real part $\varepsilon^\prime$ and (b) the imaginary part $\varepsilon^{\prime\prime}$ of the dielectric constant of Dy$_2$Ti$_2$O$_7$ for magnetic fields along the [111] directions. The data were obtained on warming after cooling under the presence of the magnetic field.}
\label{fig:EvsT}
\end{figure}

We next present the temperature dependence of the dielectric constant in Fig.~\ref{fig:EvsT-100} and~\ref{fig:EvsT}. 
In zero field, we observed a substantial reduction in the real part $\varepsilon^\prime$ below about 1.5 K, as well as a strong peak in the imaginary part $\varepsilon^{\prime\prime}$ at 1.2 K. 
These features correspond to the formation of the local 2-in, 2-out spin configurations governed by ferromagnetic nearest-neighbor interaction $J_{\mathrm{eff}}$ = 1.1 K~\cite{Bramwell}, leading to the frustrated spin-ice state with residual entropy. 
By applying a magnetic field along the [100] direction, the spin frustration is mostly released above 1.0 T.
Correspondingly, in a magnetic-field-cooling sequence above 1 T, the features in both $\varepsilon^\prime$ and $\varepsilon^{\prime\prime}$ observed at lower fields disappear completely as shown in Fig.~\ref{fig:EvsT-100}.
When the field is reduced to zero after magnetic-field-cooling at high fields, $\varepsilon^\prime$ smoothly decreases to the same value obtained in the zero-field cooling even at 0.26 K.

We next describe the dielectric response for the [111] field direction, for which details of the changes in the spin frustration are known from the specific heat and magnetization~\cite{Higashinaka2,Sakakibara}.
For the [111] field direction shown in Fig.~\ref{fig:EvsT}(a), $\varepsilon^\prime$ responds sharply to the critical end-point at (0.4 K, 1 T) of the first order transition from the Kagome-ice state with 2-in, 2-out configuration to the ``1-in, 3-out" state. 
We note that the peak position of $\varepsilon^{\prime\prime}$ shown in Fig.~\ref{fig:EvsT}(b) shifts to higher temperature by about 0.15 K from 0 to 0.37 T.

\begin{figure}[b]
    \begin{center}
\includegraphics[width=0.9\linewidth,clip]{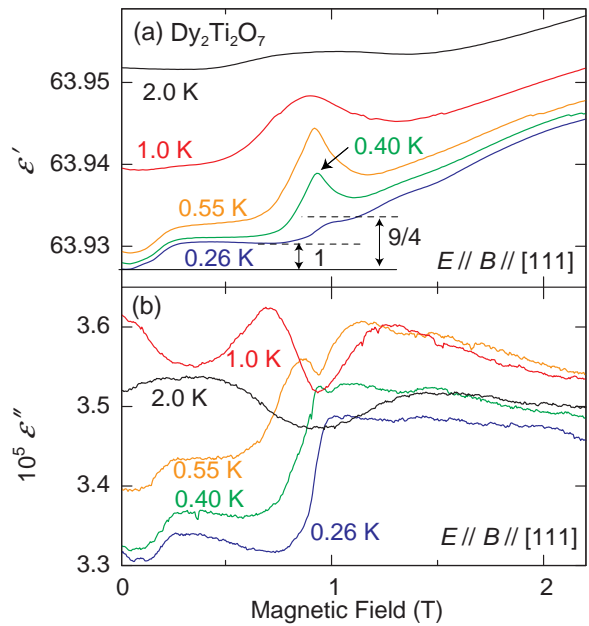}
    \end{center}
\caption{(Color online) Magnetic field dependence of (a) the real and (b) the imaginary parts of the dielectric constant of Dy$_2$Ti$_2$O$_7$ for the [111] field direction. All the data were taken on field reduction after zero-field-cooling and subsequent field application at least up to 3 T. $\varepsilon^\prime$ has been corrected for magnetostriction. See text for the dotted lines.}
\label{fig:EvsH}
\end{figure}

Fig.~\ref{fig:EvsH} shows the field dependence of the magneto-dielectric response in the [111] field direction. 
Corresponding to the magnetization plateau in the Kagome-ice state~\cite{Sakakibara}, $\varepsilon^\prime$ displays a field-independent plateau between 0.3 and 0.8 T at low temperatures as shown in Fig.~\ref{fig:EvsH}(a). 
At higher temperatures $\varepsilon^\prime$ behaves quite differently, exhibiting a peak slightly below 1 T which broadens with increasing temperature. 
The implication of this behavior becomes clearer in the dielectric loss $\varepsilon^{\prime\prime}$ in Fig.~\ref{fig:EvsH}(b). 
The $B$-dependence of $\varepsilon^{\prime\prime}$ exhibits a shoulder-like anomaly at about 0.3 T, reflecting the change from the spin-ice to Kagome-ice state.
The field-induced first-order transition below 0.4 K~\cite{Sakakibara} results in a sharp enhancement of $\varepsilon^{\prime\prime}$ centered at 0.94 T. 
At higher temperatures $\varepsilon^{\prime\prime}$ displays two strong peaks below and above 1 T. 
When plotted in the phase diagram in Fig.~\ref{fig:Phase} as closed triangles, these clearly correspond to the crossover lines from the Kagome-ice state to the ``1-in, 3-random" state and to the ``1-in, 3-out" state. 
In addition, another peak emerges in the low-field region above 1 K; the peak shifts to higher field with increasing temperature. 
This peak in $\varepsilon^{\prime\prime}$ corresponds to the crossover line between the all-random and the ``1-in, 3-random" states.

\section{Discussion}
The main features in the magneto-dielectric response may be summarized as follows: 
(1) The peaks in the dielectric loss correspond very well to the changes in the spin configuration determined from the specific heat and magnetization. 
(2) The boundary between the Kagome-ice and ``1-in, 3-random" states is precisely determined from $\varepsilon^{\prime\prime}$. 
It is interesting to note that the boundary temperature has a maximum at 1.35 $\pm$ 0.05 K for $\mu_{0}H$ = 0.5 $\pm$ 0.1 T.
While the Kagome ice state is destabilized by magnetic fields for low-$T$, high-$B$ region, it is $stabilized$ by fields for high-$T$, low-$B$ region.
If the boundary reflects the thermal activation of the least-stable spins under magnetic fields, the boundary would have a negative $dB/dT$. 
This is because the thermal energy to overcome the Zeeman energy to flip one of the spins on the Kagome plane with a field-antiparallel component of the magnetic moment is given by 2$\Delta$$E_1$=4$J_{\mathrm{eff}}$ $-$ (2/3)$g_JJ\mu_{B}B$ and decreases with increasing $B$. 
The behavior of the boundary at high $B$ is explained by this mechanism.
In fact, the above equation gives $B_{\rm c}$ = 1.0 T for the disappearance of the first excited gap, consistent with the result shown in Fig.~\ref{fig:Phase}.
On the contrary, the boundary for $B$ $<$ 0.4 T is given by $T_{{\rm b}} (B) = T_{{\rm si}} + aB$ with the spin-ice entering temperature $T_{{\rm si}} = 1.15$ K and the {\it{positive}} slope $a$ = 0.41 K/T. 
This behavior cannot be explained even if we introduce the entropy term in the free energy and indicates that this boundary is essentially a collective change in the spin configuration, rather than by a thermal activation of individual spins.

Let us now examine whether the phenomenological formula~\cite{Katsufuji1} relating the nearest-neighbor spin correlation $\langle {\bm m}_{i} \cdot {\bm m}_{j} \rangle _{\rm{nn}}$ (here $m_i=g_{\rm{J}}J_i\mu_{\rm{B}} = 10  \mu_B$ for Dy$^{3+}$) to the dielectric constant 
\begin{equation}
\varepsilon^{\prime}(T,B)=\varepsilon^{\prime}_0(T)(1 +  \alpha \langle {\bm m}_{i} \cdot {\bm m}_{j} \rangle _{\rm{nn}})
\end{equation}
is applicable to Dy$_2$Ti$_2$O$_7$. 
With magnetic fields in the [100] direction, the macroscopic degeneracy is lifted while the local 2-in, 2-out spin configuration is maintained. 
Thus $\langle {\bm m}_{i} \cdot {\bm m}_{j} \rangle _{\rm{nn}}$ should be unchanged, contrary to the observation in which the low-temperature reduction in $\varepsilon^{\prime}$ clearly ceases under the field as shown in Fig.~\ref{fig:EvsT}(a). 

To seek an alternative phenomenological description applicable to Dy$_2$Ti$_2$O$_7$, let us replace $\langle {\bm m}_{i} \cdot {\bm m}_{j} \rangle _{\rm{nn}}$ by $\langle {\bm M}_{i} \cdot {\bm M}_{j} \rangle _{\rm{nn}}$, where ${\bm M}_i$ is the total moment of the four spins on each tetrahedron, and the average is taken over the nearest-neighbor tetrahedra constituting the network of super-tetrahedra (shown in Fig.~\ref{fig:Tetra}).
Such replacement is meaningful if a TiO$_6$ octahedron, which is surrounded by Dy$_2$O tetrahedra and plays the main role in the dielectric response~\cite{Ito}, couples more strongly with a total magnetization of a Dy$_2$O tetrahedron than with an individual Dy$^{3+}$ moment.
Under zero field, the 2-in, 2-out state will yield the average $\langle {\bm M}_{i} \cdot {\bm M}_{j} \rangle _{\rm{nn}}$  = 0 since ${\bm M}_i$ randomly points in any of the six directions of $\langle 100 \rangle$. 
With increasing magnetic field along the [100] direction, all ${\bm M}_i$s become parallel to each other and $\langle {\bm M}_{i} \cdot {\bm M}_{j} \rangle _{\rm{nn}}$ = $(4 \times 1/\sqrt{3})^2$ = $16/3$, if the magnitude of the original spin moment $m_i$ is taken as unity. 
In the Kagome-ice state under the [111] field, ${\bm M}_i$ can take only three of the directions and $\langle {\bm M}_{i} \cdot {\bm M}_{j} \rangle _{\rm{nn}}$  = $16/3 \times 1/3$ = $16/9$.
With increasing field, the spin configuration changes to ``1-in, 3-out" state with $\langle {\bm M}_{i} \cdot {\bm M}_{j} \rangle _{\rm{nn}}$ = 4, giving an enhancement of $\langle {\bm M}_{i} \cdot {\bm M}_{j} \rangle _{\rm{nn}}$ by a factor of $9/4$.

Since the magnetization~\cite{Sakakibara} essentially saturates above 1.5 T, we assume for such low fields that $\varepsilon^{\prime}$($T$,$B$) is a sum of magnetic and non-magnetic contributions. By analogy to Eq. (1), we examine the relation below 1.5 T,
\begin{equation}
\varepsilon^\prime(T,B)=\beta \langle {\bm M}_{i} \cdot {\bm M}_{j} \rangle _{\rm{nn}} + A(T);
\end{equation}
thus $A(T) = \varepsilon^\prime(T,0)$.
In addition, for a given temperature $\Delta\varepsilon^\prime (T,B) = \varepsilon^\prime (T,B)- \varepsilon^\prime (T,0)$ = $\beta\langle {\bm M}_{i} \cdot {\bm M}_{j} \rangle _{\rm{nn}}$.
If we further assume that $\beta$ is constant, $\Delta\varepsilon^\prime (T,B)$ is constant for the Kagome-ice state since $\langle {\bm M}_{i} \cdot {\bm M}_{j} \rangle _{\rm{nn}}$ remains constant at $16/9$. 
Indeed, $\Delta\varepsilon^\prime(T,B)$ in the Kagome-ice state at 0.26, 0.35, and 0.40 K shown in Fig.~\ref{fig:EvsH} (a) gives the same value of 0.0033, implying that $\beta$ is about $2 \times 10^{-3}$.
Moreover, in the ``1-in, 3-out" state just above 1T, the observed value of $\Delta\varepsilon^\prime(T,B)$ is in fair agreement with the expectation of $0.0033 \times 9/4 = 0.0074$ as shown in Fig.~\ref{fig:EvsH}(a).
$\Delta\varepsilon^\prime(T,B)$ for the [100] field direction (Fig.~\ref{fig:EvsT}(a)) is also in fair quantitative agreement using the same value of $\beta$.

Let us next consider another expression,
\begin{equation}
\varepsilon^\prime(T,B)=\beta \langle \tilde{M}_{z}^{2} \rangle + \varepsilon^\prime(T,0);
\end{equation}
in which the direction ``$z$" is defined along the field, and $\tilde{M}_{z}$ is the bulk magnetization : 
($\tilde{M}_{z} = \sum_{i=1}^{N/4} M_{iz}$. $M_{iz}$ is the $z$-component of $\bm{M}_{i}$, $N$ is number of spins, and $N$/4 represents the number of tetrahedra.)
Incidentally, Eqs. 2 and 3 give the same results for the Kagome-ice state, the 2-in, 2-out polarized state under the [100] field, and the ``1-in, 3-out" polarized state. However, the two expressions give different results for a spin-ice state under weak field : In the spin-ice state in the limit of low field, $\langle {\bm M}_{i} \cdot {\bm M}_{j} \rangle _{\rm{nn}} = 0$ whereas $\frac{4}{N} \langle \tilde{M}_{z}^{2} \rangle = 16/9$. 
Furthermore, there is a great difference in the $T$-dependence.
Since $\langle {\bm M}_{i} \cdot {\bm M}_{j} \rangle _{\rm{nn}}$ is affected not only by the z-component $M_{iz}$, but also by the transverse components $M_{ix}$ and $M_{iy}$, the difference between $\langle {\bm M}_{i} \cdot {\bm M}_{j} \rangle _{\rm{nn}}$ and $\langle \tilde{M}_{z} \rangle^{2}$ increases with the increase in the thermal fluctuation, especially in zero and low field regions.~\cite{magnet}

Although neither $\langle {\bm M}_{i} \cdot {\bm M}_{j} \rangle _{\rm{nn}}$ nor $\langle \tilde{M}_{z}^{2} \rangle$ can be measured directly, $\langle \tilde{M}_{z}^{2} \rangle$ can be expressed in terms of readily measurable physical quantities (susceptibility $\chi_{z}$ and mean magnetization of a tetrahedron $\langle \tilde{M}_{z} \rangle$) as $\langle \tilde{M}_{z}^{2} \rangle = \langle \tilde{M}_{z} \rangle^{2} +  T \chi_{z}$, if thermal equilibrium is assumed. Then Eq. 3 can be written as
\begin{equation}
\varepsilon^\prime(T,B)=\beta \{ \langle \tilde{M}_{z} \rangle^{2} +  T \chi_{z}(T,B) \} + \varepsilon^\prime(T,0).
\end{equation}
The first term explains semiquantitatively the plateaus in $\varepsilon^\prime$ in the Kagome-ice and ``1-in, 3-out"states, as discssed above.

The second term $T \chi_{z}(T,B)$ is very important to explain the peak of $\varepsilon^\prime$ around 1 T (Fig.~\ref{fig:EvsH}(a)), not contained in the magnetization data~\cite{Sakakibara}.
This is the region of metamagnetic instability with large $\chi_z$.
The factor $T$ in the second term should make the contribution smaller at lower temperature, while the factor $\chi_z$ makes the peak broader at higher temperatures, just as observed.
Equation 4 thus captures the information of $\varepsilon^\prime$ which contains not only the static magnetization $\langle \tilde{M}_{z} \rangle$ but also the magnetic susceptivility $\chi_z$.

A preliminary Monte-Carlo simulation suggests that the observed peak in $\varepsilon^\prime$ at about 1 T is much greater than that expected from eq. 4 and the expression in terms of $\langle {\bm M}_{i} \cdot {\bm M}_{j} \rangle _{\rm{nn}}$ better describes the observation. 
Yet additional contribution seems to be present at higher fields: even for the case where the magnetization is saturated with increasing field, the dielectric constant keeps changing. 
The sign of the change is negative in the [100] field direction above 2 T (Fig.~\ref{fig:EvsT}(a)), while it is positive in the [111] field direction. 
This difference in sign may provide an important clue to understanding the variation of $\varepsilon^\prime$ at high fields. 

\section{Conclusion}
We have demonstrated that magneto-dielectric response can be used as a high-sensitivity and high-precision probe with fast data acquisition capabilities for the investigation of a geometrically frustrated magnet. We found that the dielectric loss $\varepsilon^{\prime\prime}$ is particularly sensitive to probe subtle changes in the spin configuration, not easily detectable by magnetostriction. 
In addition, the dielectric constant $\varepsilon^\prime$ contains information of not only the magnetization $\langle \tilde{M}_{z} \rangle$, but also of the susceptibility $\chi_{z}$.

\acknowledgements

The authors are grateful to T. Katsufuji for useful discussion and sending their data prior to publication, to Koichiro Tanaka for letting us use a sputter coating machine, to M. Gingras for useful discussion and information, and to R. Perry, K. Deguchi, K. Kitagawa, K. Ishida, H. Yaguchi, S. Nakatsuji, and S. Grigera for their technical help as well as useful advises. This work is supported by Grants-in-Aid for Scientific Research (S) from JSPS and for the 21 Century COE program ``Center for Diversity and Universality in Physics" from MEXT of Japan.


\begin{thebibliography}{99}
\bibitem{magnet}
Instead of using the bulk magnetization $\langle \tilde{M}_{z}^{2} \rangle$, one may introduce thermal average of the square of the local magnetization along the applied field $\langle {M}_{z}^{2} \rangle$.
However, this quantity neither gives zero for the low-field limit, nor can be expressed in terms of other measurable bulk quantities.
Thus we use $\langle \tilde{M}_{z}^{2} \rangle$ in the following discussion.
\end{thebibliography}
\end{document}